\documentclass[a4paper,11pt]{article}
\usepackage{url}
\usepackage{epsfig}
\usepackage{amsfonts}
\usepackage{graphicx}
\usepackage{epsfig}
\usepackage{eepic}
\usepackage{amsmath}
\usepackage{amssymb}
\usepackage{color}
\usepackage{bbm}
\usepackage{dcolumn}% Align table columns on decimal point
\usepackage{bm}% bold math
\usepackage{ulem}
\usepackage{mathrsfs}
\usepackage{bbold}
\usepackage{datetime}

\usepackage{overpic} % add text over picture
\usepackage{rotating}
\usepackage[usenames,dvipsnames]{xcolor}
\usepackage{lipsum} % for mock text
\usepackage{tikz,tikz-3dplot}
\usetikzlibrary{shapes.geometric}
\usepackage{etoolbox} % for \appto
\usepackage{extarrows}

\usepackage{jheppub}

\def\bc{\begin{center}}

\def\ec{\end{center}}
\def\be{\begin{eqnarray}}
\def\ee{\end{eqnarray}}
\definecolor{dyellow}{rgb}{1.,0.8,.0}
\definecolor{myblue}{rgb}{.1,.1,.7}
\definecolor{dcyan}{rgb}{.0,.6,.6}
\definecolor{dmagenta}{rgb}{0.6,0.0,0.6}
\definecolor{brown}{rgb}{0.6,0.2,0.}
\definecolor{darkblue}{rgb}{.0,.0,0.5}
\definecolor{darkred}{rgb}{0.75,0.0,0.0}
\definecolor{orange}{rgb}{1.,.6,.0}
\definecolor{dorange}{rgb}{0.8,.4,.0}
\definecolor{darkgreen}{rgb}{0.0,0.6,0.0}
\definecolor{purple}{rgb}{.4,.0,.4}
\definecolor{lightgrey}{rgb}{0.7, 0.7, 0.7}
\definecolor{grey}{rgb}{0.4, 0.4, 0.4}
\def\eps{\epsilon}

\usepackage[position=t, singlelinecheck=off]{subfig}
\usepackage[font=small,labelfont=bf,justification=raggedright]{caption}

\newcommand{\xdownarrow}[1]{%
  {\left\downarrow\vbox to #1{}\right.\kern-\nulldelimiterspace}
}
\newcommand{\xuparrow}[1]{%
  {\left\uparrow\vbox to #1{}\right.\kern-\nulldelimiterspace}
}

\title{Kibble-Zurek Mechanism and Current-Phase Relation in a Holographic Josephson Junction}

\author[a,b]{Zhi-Hong Li,}
\author[c,b]{Huai-Fan Li}
\author[d,e]{and Hai-Qing Zhang}
\affiliation[a]{Department of Physics, Shanxi Datong University, Datong 037009, China}
\affiliation[b]{Institute of Theoretical Physics, Shanxi Datong University, Datong 037009, China}
\affiliation[c]{College of General Education, Shanxi College of Technology, Shuozhou 036000, China}
\affiliation[d]{Center for Gravitational Physics, Department of Space Science, Beihang University,
Beijing 100191, China}
\affiliation[e]{Peng Huanwu Collaborative Center for Research and Education, Beihang University, Beijing
100191, China}

\emailAdd{lizhihong@buaa.edu.cn}
\emailAdd{huaifan999@163.com}
\emailAdd{hqzhang@buaa.edu.cn}
\abstract{We present a comprehensive study of the current-phase relation of the Josephson junction in a holographic superfluid ring, realized from the stochastic and non-equilibrium dynamics according to the Kibble-Zurek mechanism. By employing a spatially modulated charge density to engineer a weak link, the current-phase relation is investigated in a range of geometric and thermodynamic parameters. The seminal sinusoidal relation between the current and the phase emerges periodically due to the compact shape of the geometry. We also identify the relations between the critical current and the geometric parameters of the junction: the width, steepness and depth. Furthermore, we demonstrate that the critical current exhibits a characteristic exponential decaying against the final temperature, reflecting the thermal degradation of the order parameter in a strong-coupling regime. Our results establish a robust framework for holographic Josephson devices, offering experimentally testable predictions for the non-equilibrium dynamics of high-$T_c$ superconductors.}
\begin{document}
\maketitle
\flushbottom

%\tableofcontents

\section{Introduction}

The current-phase relation of a Josephson junction - the dependence of the supercurrent on the gauge-invariant phase difference across a weak link - stands as one of the most incisive diagnostics in modern quantum matter research \cite{Josephson:1962}. Experimentally, such weak links are engineered via three primary architectures: the superconductor-insulator-superconductor (SIS) junction, the superconductor-normal-superconductor (SNS) junction, and the superconductor-superconductor-superconductor (SSS) junction \cite{Likharev1979}. Theoretically, the current-phase relation encodes critical information about the pairing symmetry, quantum tunneling, and topological properties of the superconducting state \cite{Golubov2004}. While conventional Josephson junctions obey the seminal sinusoidal law such as $J_x \propto \sin\gamma$, where $J_x$ stands for the supercurrent and $\gamma$ is the phase difference,  deviations from this behavior - such as skewed or multivalued profiles - have been observed in high-$T_c$ and topological superconductors, implying an exotic underlying physics \cite{Sun1994,Buzdin2005}. However, a systematic understanding of how geometric structures, thermal fluctuations, and topological order jointly govern $J_x(\gamma)$ in strongly-correlated systems remains lacking.  

Holographic Josephson junctions (HJJs) offer a unique platform to address this gap. At the heart of this framework lies the AdS/CFT correspondence (or gauge-gravity duality) \cite{Maldacena:1997re,Witten:1998qj,Gubser:1998bc}, a conjecture which has now been recognized as a powerful tool to model the unconventional superconductivity beyond the BCS paradigm \cite{Hartnoll:2008vx}. Recent progress has established the holographic Josephson junction as a controllable model in which the strong-coupling physics of superconductivity can be probed well beyond the BCS mechanism. The benchmark construction in \cite{Horowitz:2011dz} demonstrated that a spatially varying chemical-potential profile in an AdS–Schwarzschild bulk naturally engineers a weak link on the boundary, yielding the seminal sinusoidal current–phase relation $J_x \propto \sin\gamma$. Besides, they also uncovered the exponential decaying between the critical current $J_{\text{max}}$ and the temperature as well as the width of the junction. This work was subsequently lifted to $(3+1)$-dimensional boundary theories \cite{Wang:2011rva}, recast in the language of explicitly inhomogeneous holographic condensates, and extended beyond the s-wave channel to p-wave configurations \cite{Wang:2011ri}.   Alternative models are also studied ranging from AdS/BCFT defect descriptions and Josephson networks \cite{Kiritsis:2011zq}, to double-trace deformations of the boundary CFT \cite{Domokos:2012rj}, anisotropic/Lifshitz backgrounds and backreacting geometries \cite{Wang:2012yj,Takeuchi:2013kra,Wang:2016hjw,Li:2014xia,Cai:2013sua,Liu:2015zca,Takeuchi:2021kjv}. Nevertheless, the majority of these literatures remain within the static, equilibrium setup.  By contrast, using nonequilibrium methods to unveil how a junction relaxes into those states when driven by a quench, how phase equilibrates, and whether the equilibrium critical current $J_{\text{max}}$ scaling still survives by finite-rate dynamics remains unexplored. It is precisely this gap that motivates us of the present work.

In this work, we perform a fully dynamical analysis of the current-phase relation and critical current scaling within the framework of the spatial one-dimensional holographic SSS Josephson junction. Unlike previous static studies, we solve the full time-dependent Einstein-Maxwell-scalar equations to track the relaxation of the superfluid order parameter and phase after a quench. This enables us to extract the current-phase relation $J_x(\gamma)$ in the final equilibrium state and investigate how the critical current $J_{\text{max}}$ is affected by the parameters of the Josephson junction, such as the width, steepness and depth. To capture the genuine nonequilibrium dynamics of Josephson junctions, we must go beyond the adiabatic approximation and consider the Kibble-Zurek mechanism (KZM) as a starting point \cite{Kibble:1976sj,Kibble:1980mv,Zurek:1985qw}. KZM provides a universal framework for understanding how topological defects (in our case it is the winding number of the superfluid phase) emerge when a system is driven through a continuous phase transition. Specifically, as the system approaches the critical temperature $T_c$ from above, critical slowing down causes the relaxation time to diverge. Once the quench rate exceeds the system's ability to respond adiabatically, the order parameter dynamics effectively freeze, locking in topological defects in the broken-symmetry phase. The universality of the KZM has been firmly established through its successful application across a diverse array of experimental platforms, ranging from superfluid helium and liquid crystals to ultra-cold atomic Bose-Einstein condensates \cite{Chuang:1991zz,Ruutu:1995qz,Carmi:2000zz}. Within the framework of gauge-gravity duality, KZM has emerged as a powerful tool to probe the non-equilibrium dynamics of strongly correlated systems. Early holographic studies laid the groundwork by analyzing phase quenches in (1+1)-dimensional rings \cite{Sonner:2014tca} and demonstrating the spontaneous formation of vortex defects in (2+1)-dimensional superfluids \cite{Chesler:2014gya}. It was later significantly advanced by studies of (2+1)-dimensional holographic superconductors, which revealed that the characteristic topological remnants of a quench are fluxons -- quantized magnetic fluxes trapped within the cores of vortices \cite{Zeng:2019yhi}. More recent investigations have further generalized these principles, exploring the impact of inhomogeneities, varied quench protocols, and complex geometries such as holographic rings and multi-component systems \cite{Li:2019oyz,Xia:2020cjl,delCampo:2021rak,Li:2021iph,Li:2021dwp,Li:2021mtd,Xia:2021xap,Li:2021jqk,delCampo:2022lqd,Li:2022tab,li:2024twf}.
%{\blue{ Although the current-phase relations characterize the equilibrium properties of the junction, the KZM protocol provides a unique non-equilibrium pathway to probabilistically prepare excited winding states. The novelty here is not in the static observable itself, but in demonstrating that the defect density—and thus the population of different superconducting sectors—follows universal scaling laws determined by the quench dynamics.}}

In this work, we employ a linear temperature quench to drive the system through a continuous (second-order) phase transition from the normal to the superconducting state. We focus on the dependence of the relation between the geometry of the junction: the length $\mathcal{L}$, the depth $\epsilon$ and the steepness $\sigma$. The dependence of the critical current between the final temperature $T_f$ is also investigated. Except the steepness, we find an exponential relation between the critical current and the above parameters.  These studies allow us to quantify how the interplay between the quench protocol and the junction geometry reshapes the universal scaling laws of $J_{\text{max}}$.

We begin in Sec.~\ref{two} by constructing the holographic dual of a superfluid ring interrupted by a weak link. With this framework in place, Sec.~\ref{three} presents our core findings: a detailed numerical analysis of the current-phase relation and critical current scaling in the holographic SSS junction. We conclude in Sec.~\ref{four} with a summary of our key observations and discussions.

\section{Holographic setup}\label{two}
\subsection{Einstein-Maxwell-scalar model}
Our starting point is the Einstein-Maxwell-scalar system, where a $U(1)$ gauge field and a charged complex scalar field are minimally coupled to gravity in a (3+1)-dimensional spacetime. The dynamics of the system are governed by the Abelian-Higgs action \cite{Hartnoll:2008vx}, 
\begin{eqnarray}\label{density}
S=\int d^4x\sqrt{-g} \left( -\frac{1}{4} F_{\mu \nu} F^{\mu \nu} - |\nabla\Psi -iA \Psi|^2 - m^2 |\Psi|^2\right).
\end{eqnarray}
in which  $F_{\mu\nu}=\partial_\mu A_\nu-\partial_\nu A_\mu$ denotes the field strength tensor of the $U(1)$ gauge field $A_\mu$, and the complex scalar field is parameterized as $\Psi=|\Psi|e^{i\theta}$.
In the probe limit, the bulk equations of motion read:
\begin{eqnarray}\label{eomofwhole}
D_\mu D^\mu\Psi=m^2\Psi, \qquad \qquad \nabla_\mu F^{\mu\nu}=j^\nu, 
\end{eqnarray}
where the current is $j^\nu=i(\Psi^* D^\nu\Psi-c.c.)$ in which $c.c.$ represents {\it complex conjugation}. 

We employ the Eddington-Finkelstein coordinates in an AdS$_4$ planar black hole \cite{Chesler:2013lia} to capture the full time-dependent evolution of the system,
\begin{eqnarray}
ds^2 = \frac{1}{z^2} \left(-f(z) dt^2 - 2dtdz + dx^2 +dy^2 \right),
\end{eqnarray}
where $z$ is the radial coordinate and $f(z) = 1 - z^3$ (We have set the AdS radius $l=1$ and the horizon $z_h=1$.). The infinite boundary resides at $z = 0$, corresponding to a dual field theory with temperature $T=3/(4\pi)$.
To simulate a one-dimensional ring on the boundary, we compactify the $x$-direction with periodic boundary conditions and assume homogeneity along $y$-direction. This leads to the consistent field configuration: $\Psi = \Psi(t,z,x), A_{t} = A_{t}(t,z,x), A_x=A_x(t,z,x)$ and $A_{y,z} = 0$.  Under this ansatz, the dynamical equations Eq.\eqref{eomofwhole} reduce to:
 \begin{eqnarray}
\label{eompsi}
\partial_t \partial_z \psi  - \frac12 \big[( i \partial_z A_t -z-i \partial_x A_x - A_x^2 )\psi  + (\partial_z f +2 i A_t )\partial_z \psi  + f \partial_z^2 \psi - 2 i A_x \partial_x \psi+ \partial_x^2 \psi  \big] = 0;~~~~&
\\
\label{eom2}
\partial_t \partial_z A_t - \partial_x( \partial_x A_t  + f \partial_z  A_x  - \partial_t  A_x) 
+ 2 A_t |\psi|^2 + i \big[ f (\psi \partial_z \psi^*-\psi^* \partial_z \psi) -(\psi \partial_t \psi^*-\psi^* \partial_t \psi )\big] = 0;~~~~&
\\
\label{eom3}
\partial_t \partial_z A_x - \frac12 \big[ \partial_z (\partial_x A_t + f \partial_z A_x) + i (\psi \partial_x \psi^*-\psi^* \partial_x \psi ) \big] +  A_x |\psi|^2 = 0;~~~~&
\\
\label{eom1}
\partial_z^2A_t-\partial_z \partial_x A_x  + i (\psi \partial_z \psi^*-\psi^* \partial_z \psi) = 0.~~~~&
\end{eqnarray}
where we have scaled $\psi=\Psi/z$.
The above four equations are not independent since they satisfy the following constraint equation,
\begin{eqnarray}
\frac{d}{dz}\text{Eq.\eqref{eom2}}-2\frac{d}{dx}\text{Eq.\eqref{eom3}}+\frac{d}{dt}\text{Eq.\eqref{eom1}}- 2i\left(\text{Eq.\eqref{eompsi}}\times\psi^*-c.c.\right)=0.
\end{eqnarray}
Therefore, we have three independent equations with three fields $\psi, A_t$ and $A_x$. Note that $\psi=\Psi/z$ is a complex field, meaning that there are four independent real fields with four independent real equations. It further implies that our choice of the gauge $A_z=A_y=0$ is feasible for the system.

\subsection{Boundary conditions and numerical schemes}
In the asymptotic region of the AdS$_4$ boundary, the bulk scalar field $\Psi$ admits the following expansion:\begin{eqnarray}
\Psi\sim \Psi_0 z^{\Delta_-}+\Psi_1 z^{\Delta_+}+\dots.
\end{eqnarray}
where the exponents $\Delta_{\pm} = \frac{3}{2} \pm \sqrt{\frac{9}{4} + m^2}$ correspond to the conformal dimensions of the dual scalar operator $\mathcal{O}$ in the boundary field theory. 
To ensure the theory resides within the Breitenlohner-Freedman (BF) stability bound ($m^2 \ge m^2_{\text{BF}} = -9/4$), we fix the scalar mass squared to be $m^2 = -2$. This specific choice yields $\Delta_{-} = 1$ and $\Delta_{+} = 2$. 
Thus, close to the conformal boundary ($z \to 0$), the scalar field $\Psi(t, z, x)$ admits the following asymptotic form:
\begin{eqnarray}
\Psi= z\left(\Psi_0+\Psi_1 z+\dots\right),
\end{eqnarray}
Without loss of generality, we adopt the standard quantization scheme where the coefficient of the subleading term $\Psi_1$ corresponds to the vacuum expectation value $\langle \mathcal{O} \rangle$, while the leading term $\Psi_0$ acts as the source. In order to be consistent with the standard quantization prescription for holographic superconductors \cite{Hartnoll:2008vx}, we impose a vanishing source condition at the AdS boundary ($z \to 0$), i.e. $\Psi_0 = 0$. This boundary condition is crucial as it ensures the absence of external fields that would explicitly break the $U(1)$ symmetry, thereby guaranteeing that any condensation of the scalar field corresponds to genuine spontaneous symmetry breaking. 

Correspondingly, the asymptotic structure of the bulk gauge field $A_\mu$ near the AdS boundary is given by
\begin{eqnarray}\label{eomat}
A_\mu&\sim& a_\mu+b_\mu z+\dots, \qquad {\text{where}~\mu=t,x}
\end{eqnarray}
in which the leading coefficient $a_t$ corresponds to the boundary chemical potential $\mu$ \footnote{Please do not confuse it with the subscript $\mu$ in $A_\mu$.} while $a_x$ corresponds to the potentials of the spatial component of gauge fields; In contrast, the subleading term $b_t$ encodes the charge density $\rho$ while $b_x$ corresponds to the current $J_x$, respectively.

At the black hole event horizon $z = z_h$, we impose the regularity condition required for a finite-temperature field theory. We fix the temporal gauge by demanding $A_t(z_h) = 0$, which eliminates unphysical degrees of freedom associated with the gauge field's residual symmetry at the horizon. Simultaneously, we enforce that all other bulk fields -- including the scalar $\Psi$ and the spatial components of the gauge field $A_x$ -- remain finite at the event horizon.

Guided by the seminal work in holographic superconductivity \cite{Hartnoll:2008vx}, we recall that the dynamics of the boundary field theory are governed by a competition between the temperature $T$ and the charge density $\rho$. In this framework, increasing the charge density $\rho$ is physically analogous to lowering the temperature $T$, driving the system deeper into the ordered phase. To see this quantitatively, a simple dimensional analysis suffices: in natural units ($\hbar = c = k_B = 1$), the black hole temperature $T$ carries mass dimension $[T]=1$, while the charge density $\rho$  carries $[ \rho ] = 2$. Consequently, the ratio $T/\sqrt{\rho}$ is a dimensionless control parameter that uniquely characterizes the distance from the critical point. We then implement a linear temperature quench following the KZM \cite{Kibble:1976sj,Kibble:1980mv,Zurek:1985qw}. The quench protocol is defined by $T(t)/T_c = 1 - t/\tau_Q$, where $\tau_Q$ denotes the quench rate. Correspondingly, the charge density $\rho$ is driven as
\be
\rho(t) = \frac{\rho_c}{(1 - t/\tau_Q)^2},
\ee
with the critical charge density $\rho_c \approx 4.06$ for the homogeneous static holographic superconductor. Quenching from a higher initial temperature $T_i > T_c$ to a final lower temperature $T_f<T_c$, the system evolves from the normal metallic phase into a superconducting state.

In our work, the numerical evolution is performed by using a fourth-order Runge-Kutta method with a fixed time step of $\Delta t = 0.1$. Spatial discretization is handled through a hybrid spectral approach: Chebyshev pseudo-spectral methods with 21 grid points resolve the radial AdS direction $z$, while Fourier decomposition with 201 grid points discretize the periodic $x$-direction, in which $x \sim x + L$ defines the ring circumference $L$.
To prepare the initial state, we first thermalize the system completely. This involves introducing Gaussian white noise $\xi(x_i, t)$ into bulk fields, characterized by $\langle \xi(x_i, t) \rangle = 0$ and $\langle \xi(x_i, t) \xi(x_j, t') \rangle = h \delta(t - t') \delta(x_i - x_j)$, with a small amplitude $h = 10^{-3}$. \footnote{ Our numerical results are robust to the values of the amplitude of the Gaussian white noise $h$. In numerics, we have checked that $h\in[10^{-8},10^{-2}]$ will produce similar results within errors.} This ensures the system starts from a well-defined thermal configuration. While the AdS black hole background encodes the initial thermal state, the introduced Gaussian white noise serves a distinct purpose: it provides the necessary microscopic perturbations to trigger spontaneous symmetry breaking during the rapid quench. Without these stochastic seeds, the homogeneous initial state will remain in an unstable symmetric phase due to numerical symmetries. Consequently, there will be no defects turned out.

In order to construct the holographic Josephson junction, we introduce a dynamically and spatially dependent perturbation to the boundary charge density $\rho(t,x)$ at the AdS boundary, thereby breaking the homogeneity of the superconducting state.
This deformation effectively creates a weak link in the superfluid ring, partitioning the system into a superconducting state connected by a weak link. The geometry is chosen such that the junction extends along the $x$-direction, while translational invariance is preserved along the transverse $y$-coordinate. The profile of the charge density $\rho(t,x)$ takes the form:
\begin{eqnarray}
\rho(t,x)=\frac{\rho_c}{\left(1-t/\tau_Q\right)^{2}}\left\{1-\frac{1-\epsilon}{2 \tanh(\frac{\mathcal{L}}{2\sigma})}\left[\tanh\left(\frac{x+\frac{\mathcal{L}}{2}}{\sigma}\right)-\tanh\left(\frac{x-\frac{\mathcal{L}}{2}}{\sigma}\right)\right]\right\},
\end{eqnarray}
where $\rho_c$ denotes the critical charge density of the homogeneous, static holographic superconductor, marking the threshold for the onset of the condensed phase. The junction geometry is controlled by three key parameters: the width $\mathcal{L}$, which sets the spatial extent of the weak link; the steepness $\sigma$, which governs the sharpness of the interface between the superconducting reservoir and the link; and the depth $\epsilon$, which quantifies the degree of suppression of the charge density within the junction. \footnote{Please note that $\eps\in[0,1]$ and smaller $\eps$ corresponds to a deeper junction.} In total, these parameters allow us to systematically tune the barrier transparency and explore the crossover from short to long junction regimes.

\subsection{Winding numbers and phase difference}
\begin{figure}[t]
\centering
\includegraphics[trim=0.cm 0.cm 0cm 0cm, clip=true, scale=0.35]{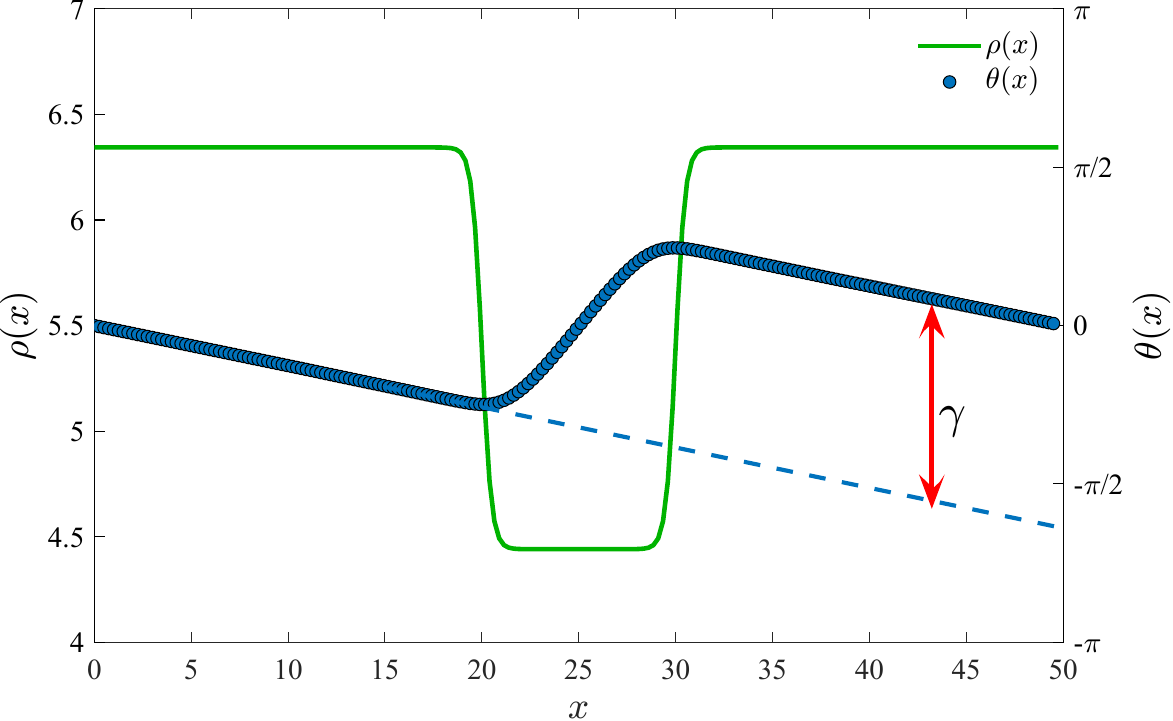}
\put(-205,120){(a)}~~~
\includegraphics[trim=0.cm 0.cm 0cm 0cm, clip=true, scale=0.35]{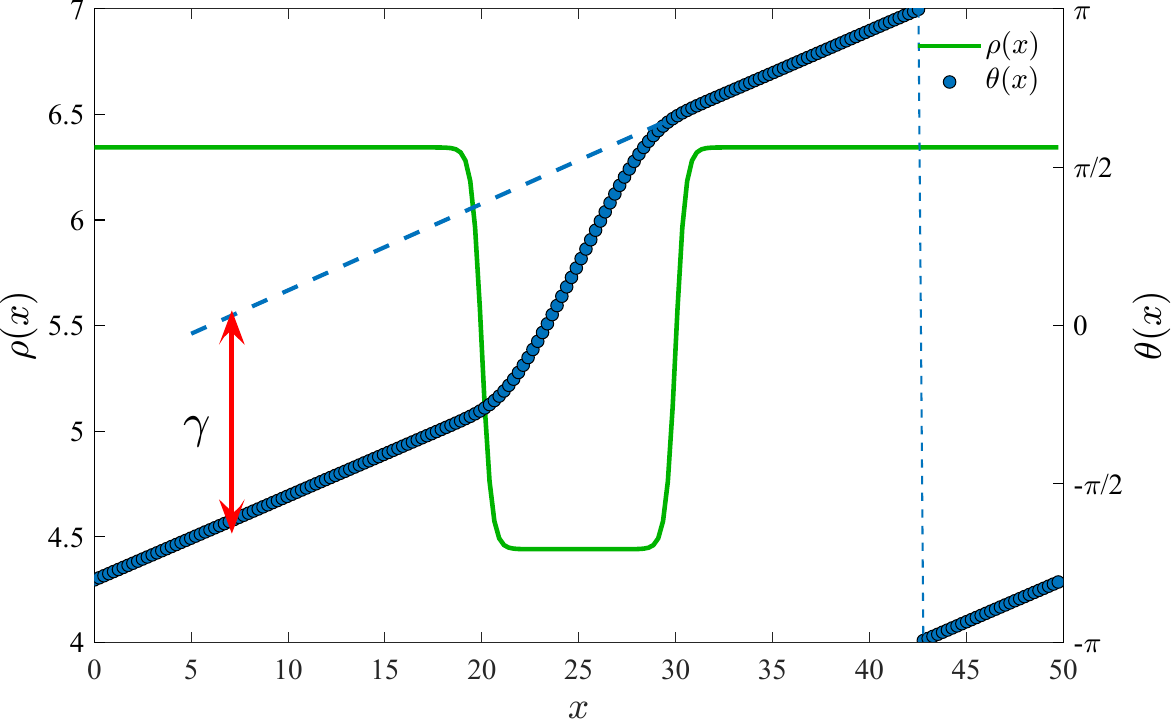}
\put(-205,120){(b)}
\caption{Holographic superfluid steady states on a compact ring with a weak link (the junction is the green sinking part). The charge density $\rho(x)$ (green solid line) and phase $\theta(x)$ (blue dotted line) are shown for two different winding numbers: $W=0$ (panel (a)) and $W=1$ (panel (b)), with fixed parameters \{$\mathcal{L}=10$, $T_f=0.8T_c$, $\sigma=0.5$, $\epsilon=0.7$\} in the final equilibrium state. The phase difference $\gamma$ across the weak link is marked by the red arrows. }
\label{p0}
\end{figure}

According to the KZM \cite{Kibble:1976sj,Kibble:1980mv,Zurek:1985qw}, a system driven through a continuous phase transition exhibits critical slowing down, causing it to fall out of equilibrium and spontaneously form topological defects as it enters the broken-symmetry phase. In the context of our holographic superfluid ring with a weak link, these topological defects manifest as quantized phase windings encircling the ring. The relevant topological invariant is the winding number $W$, which measures the net rotation of the superfluid order parameter's phase $\theta(x)$ around the compact spatial dimension. Formally, $W$ is defined via the contour integral over the ring's circumference $L$:
\begin{equation}\label{eqw}
W = \oint_L d\theta/2\pi = \frac{1}{2\pi} \int_0^L \partial_x \theta \, dx \in \mathbb{Z}.
\end{equation}
In our numerical implementation, we fix the ring's circumference as $L = 50$ (corresponding to the periodic domain $x \in [0, 50]$). Following the KZM framework \cite{Das:2011cx,Sonner:2014tca}, we initiate the dynamics by rapidly quenching the system from an initial high-temperature normal state ($T_i = 1.4\,T_c$) to a final lower temperature  $(T_f<T_c)$ superconducting state. The system is then held at $T_f$ until the non-equilibrium dynamics ceases and settles into a final equilibrium state. During this relaxation process, the phase $\theta(x)$ evolves and relaxes into straight lines with different gradients in the regime of the weak link (see Fig.\ref{p0}), providing a direct signature of the KZM at work in the presence of the Josephson junction.

To fully characterize the post-quench state, we must distinguish between the two possible orientations of the phase. We therefore define a winding number $W$ with signs. Specifically, $W = +n$ (with $n \geq 0$ and $n \in \mathbb{Z}$) corresponds to the scenario where the phase $\theta(x)$ winds from $-\pi$ to $+\pi$ exactly $n$ times as one traverses the ring along the positive $x$-direction. Conversely, $W = -n$ is defined analogously for a phase that winds $n$ times in the opposite direction.
This sign convention is physically significant: in the presence of the weak link, the sign of $W$ determines the direction of the induced superfluid velocity. As the system relaxes after the quench, these integer-valued winding configurations become frozen topological remnants, reflecting the stochastic distribution of phase gradients predicted by the KZM.

The presence of the weak link renders the equilibrium phase configurations inherently piecewise smooth. Because the numerical scheme confines the phase to the range $\theta \in [-\pi, \pi]$, the apparent discontinuities (vertical dashed lines in Fig.~\ref{p0}(b)) represent the unavoidable $2\pi$ phase slips, which are artifacts of the branch cut rather than physical singularities. When the system is rapidly quenched across the critical point $T_c$, the KZM governs the non-equilibrium dynamics, leading to the stochastic formation of topological defects that manifest as integer winding numbers in the final state. Fig.~\ref{p0} elucidates this interplay between topology and spatial inhomogeneity by contrasting two steady-state regimes: $W=0$ (panel(a)) and $W=1$ (panel(b)). In both panels, dual vertical axes denote the local charge density $\rho(x)$ (green solid line) and the unwrapped superfluid phase $\theta(x)$ (blue dotted line).  

In our holographic model, the phase difference $\gamma$ across the weak link is defined as shown in the Fig.\ref{p0} (a) and (b), in which the red arrows indicate the phase difference $\gamma$. One should distinguish the differences of our model to the previous holographic model of the Josephson junction. In previous work \cite{Horowitz:2011dz,Wang:2011ri,Wang:2012yj}, people worked in the static case. Therefore, they can use the gauge-invariant quantity to solve the system without the explicit appearance of the phase $\theta$ in the equations. However, in our case we solve the system by investigating the dynamics of the order parameter and retaining the phase $\theta$. Therefore,  we can read off the phase difference $\gamma$ directly from the final equilibrium state of the phase. For instance of the $W=0$ case in Fig.\ref{p0}(a), if there is no weak link the phase difference should be zero since the phase at equilibrium is a horizontal straight line.  However, because of the existence of the weak link, the phase within the link are different from those outside of the link. From the panel (a) we can see that they have different gradients. Therefore, we can define the phase difference $\gamma$ across the weak link as the vertical discrepancy of the phases in the two sides \cite{Piazza:2009cib}. For the non-trivial winding number $W\neq0$ case (such as Fig.\ref{p0}(b)), if there is no weak link in the middle, the phase difference would be $2\pi W$ of the system. Therefore, as the weak link exists we should further add the phase difference $\gamma$ into it, that is $2\pi W+\gamma$ as the genuine phase difference.

\section{Current–phase relation }\label{three}
In this section, we will move beyond the static topological classification and investigate the current-phase relation of the Josephson junctions. We systematically investigate the interplay between the supercurrent and the junction geometry as well as the post-quench thermal state. Specifically, we examine how the critical current $J_{\text {max}}$ and the current-phase relation $J_x(\gamma)$ curve respond to the four pivotal controlling parameters: the width $\mathcal{L}$, steepness $\sigma$, and depth $\epsilon$ of the weak link, as well as the final quench temperature $T_f$.\footnote{Please note that we have performed 100 independent stochastic realizations for each set of parameters. This ensures sufficient statistics to average out possible fluctuations. }

\subsection{The role of junction width $\mathcal{L}$}\label{subsec:width}

\begin{figure}[h]
\centering
\includegraphics[trim=0.cm 0cm 0cm 0cm, clip=true, scale=0.385]{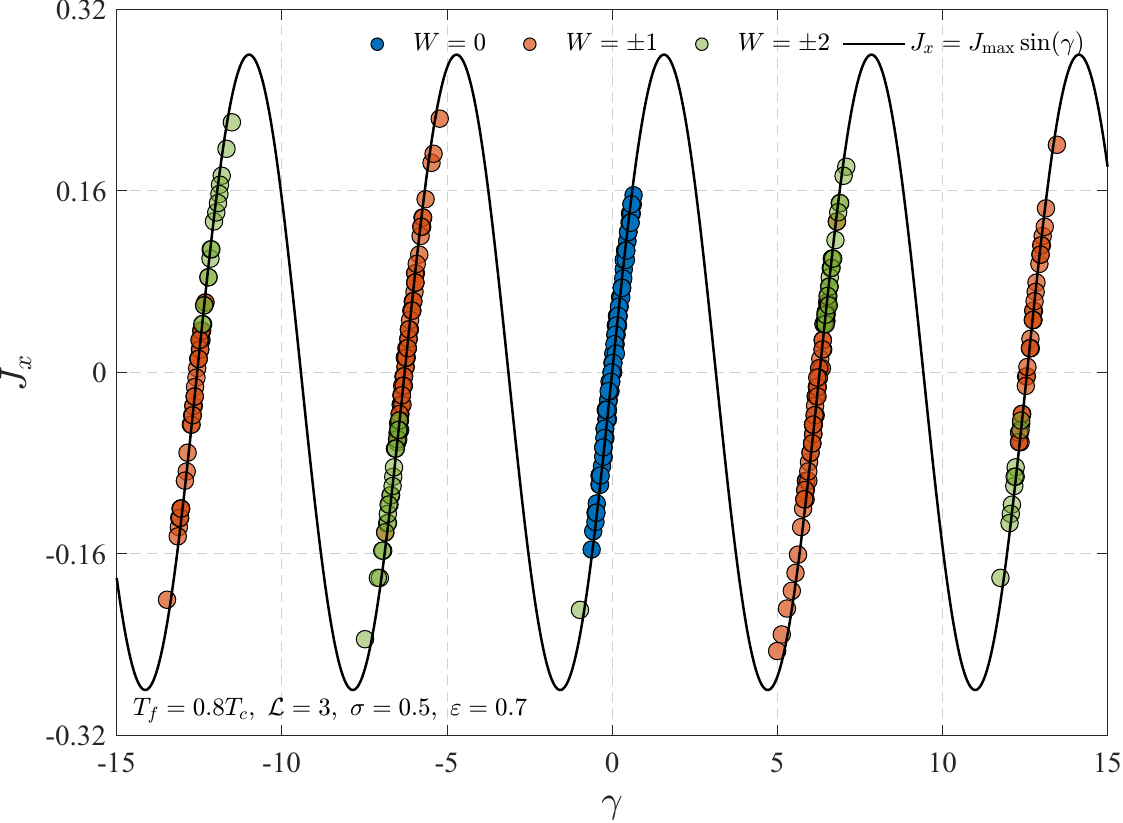}
\put(-215,150){(a)}~~~
\includegraphics[trim=0.cm 0cm 0cm 0cm, clip=true, scale=0.385]{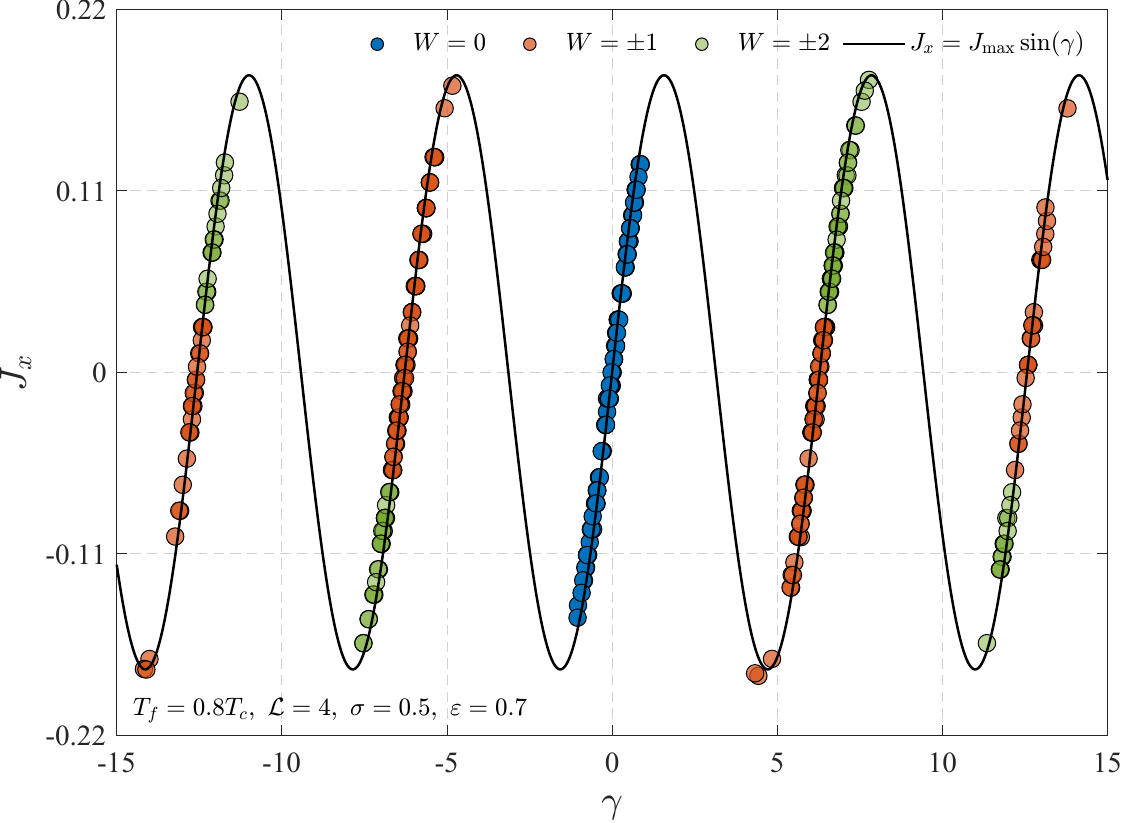}
\put(-215,150){(b)} \\
\includegraphics[trim=-0.15cm 0cm 0cm -1cm, clip=true, scale=0.36]{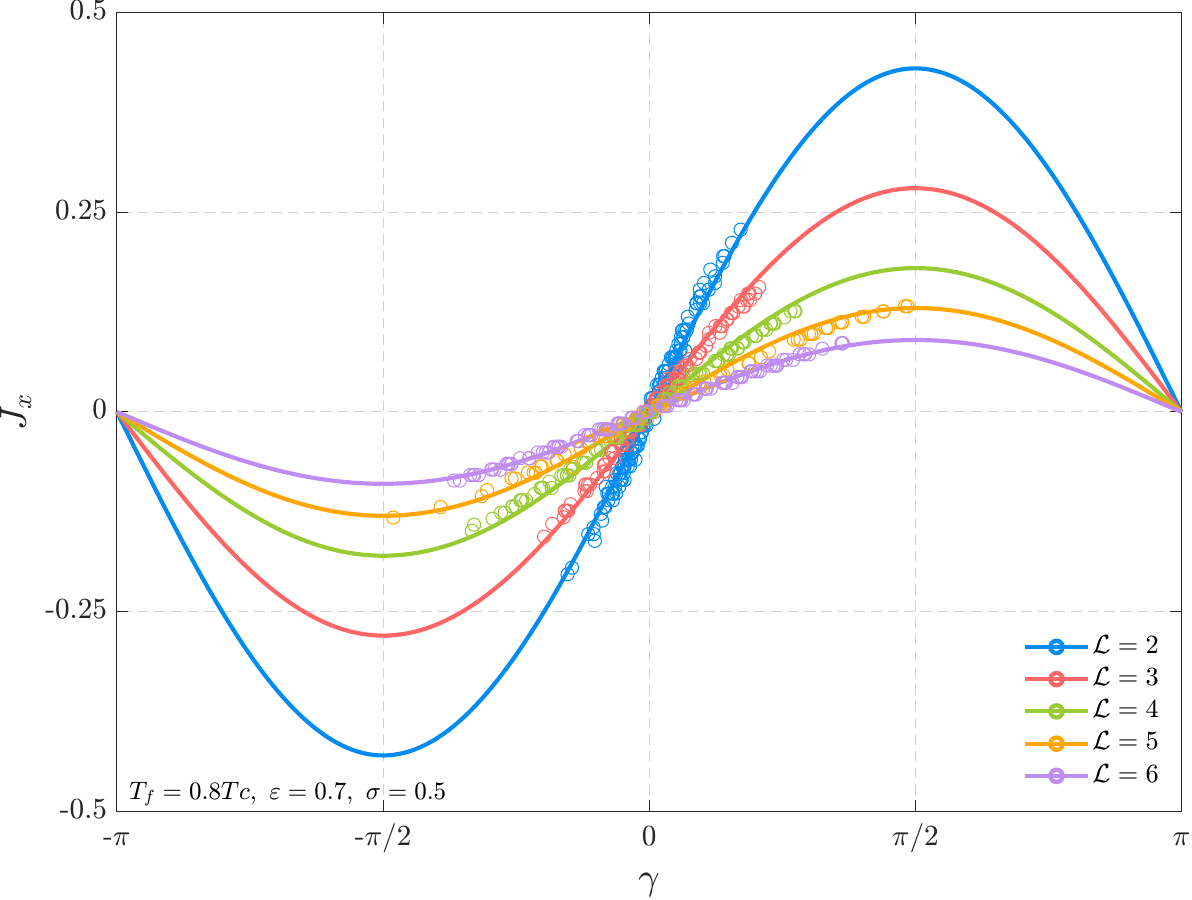}
\put(-215,150){(c)}~~~~
\includegraphics[trim=-0.cm -0.5cm 0cm 0cm, clip=true, scale=0.38]{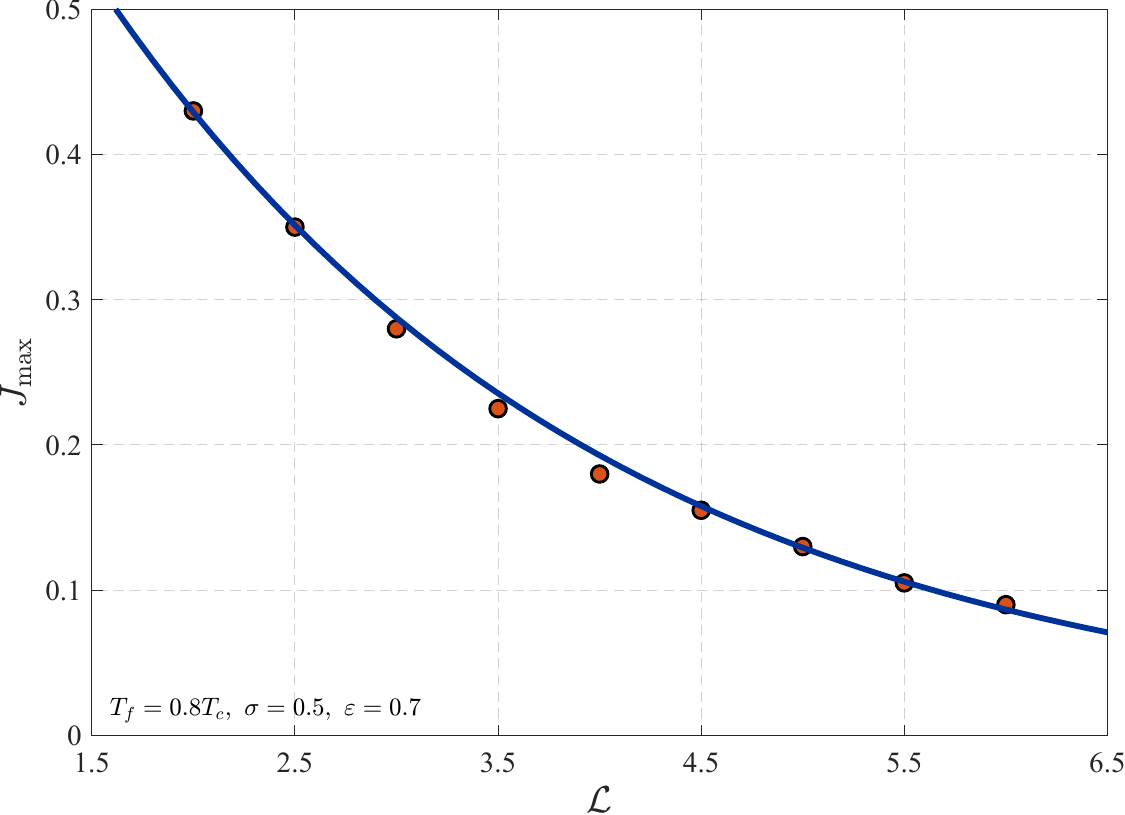}
\put(-215,150){(d)}
\put(-130,140){{\bf --} \footnotesize{$J_{\max} \approx 0.955 \, \exp\bigl(-0.40\,\mathcal{L}\bigr)$}}
\caption{The current-phase relation $J_x(\gamma)$ is plotted as a function of the phase difference $\gamma$ across the junction, exhibiting strong dependence on the junction width $\mathcal{L}$. (a) and (b), Current-phase relation with $\mathcal{L}=3$ and $\mathcal{L}=4$ for quantized winding numbers $W=\{0, \pm 1, \pm2\}$ (colored circles); The black solid lines are the fitted sinusoidal relations. (c) Various current-phase relations for the winding number $W=0$ with different junction widths $\mathcal{L}$. (d) Relation of the maximum supercurrent $J_{\text{max}}$ against $\mathcal{L}$, revealing an exponential decay of the critical current with junction width. In all panels, other parameters are fixed as $\sigma$=0.5, $\epsilon$=0.7, and $T_f$ =0.8$T_c$. }
\label{p1}
\end{figure}

 We commence our analysis by exploring the dependence of the current-phase relation on the junction width $\mathcal{L}$. Fig.\ref{p1} studies the current-phase relation $J_x(\gamma)$ for a holographic Josephson junction quenched to a final temperature of $T_f=0.8T_c$, with fixed potential depth $\epsilon=0.7$ and steepness $\sigma=0.5$. Panels (a) and (b) contrast the dynamical response for junction widths $\mathcal{L}=3$ and $\mathcal{L}=4$, respectively, with the phase difference $\gamma$ ranging from $[-15,15]$. Within this parameter space, the system supports discrete quantized winding states $W=\{0, \pm 1, \pm 2\}$, visualized as clusters of blue, red, and green circles, respectively. Crucially, these raw data points can be fitted to the celebrated sinusoidal relation (solid black lines) of the current-phase relation,
\be \label{CPR}
J_x=J_{\rm{max}}\sin(\gamma),
\ee
 confirming the robustness of the Josephson effect. 
 
 It is worth noting that for the $W=0$ states, the numerical data (blue points) are mostly distributed in the vicinity of $\gamma=0$. On the contrary, for $W\neq0$ states (green and red points), they are distributed away from the vicinity of $\gamma=0$. Besides, the numerical data will distribute on the increasing part of the sinusoidal relations, they seldom distribute on the decreasing part of the relation \eqref{CPR}. We attribute this interesting behavior to the requirement of the lower free energy. Since for larger phase difference, see for instance of the Fig.\ref{p0}, it implies higher gradient of the phase $\theta(x)$. Subsequently, greater $\nabla\theta$ will induce higher free energy since the free energy density $\mathcal{F}\propto |\nabla\theta|^2$ \cite{tinkham}. Thus, the phase difference will mostly concentrate near the vicinity of $\gamma=\pm2n\pi$ where $n=0, 1,  2 \cdots$.  \footnote{But in later sections we will see exceptions that very few data will also reside beyond the vicinity of $\gamma=\pm2n\pi$. }

For the increasing width of junction, the relationship between the current $J_x$ and the phase difference $\gamma$ are shown in Fig.\ref{p1} (c). In this panel (c), we focus ourselves primarily on the case of  $W = 0$. The phase difference $\gamma$ with various junction widths $\mathcal{L}=\{2,3,4,5,6\}$ are denoted in distinct colors. All profiles exhibit a symmetric, quasi-sinusoidal current-phase relation with $J_x$ vanishing at $\gamma=0$ and peaking near $\gamma=\pm \pi/2$. The colored solid lines are the fitting sinusoidal results such that
\begin{subequations}
\begin{align}
J_x &\approx 0.43 \sin (\gamma),  ~~~~~~~\text{for}~~~\mathcal{L} = 2, \\
J_x &\approx 0.28 \sin (\gamma),  ~~~~~~~\text{for}~~~\mathcal{L} = 3, \\
J_x &\approx 0.18 \sin (\gamma), ~~~~~~~ \text{for}~~~\mathcal{L} = 4, \\
J_x &\approx 0.13 \sin (\gamma), ~~~~~~~ \text{for}~~~\mathcal{L} = 5,\\
J_x &\approx 0.09 \sin (\gamma), ~~~~~~~\text{for}~~~\mathcal{L} = 6.
\end{align}
\end{subequations}

From panel (c), the relationship between the maximal current $J_{\text{max}}$ and the width of the junction $\mathcal{L}$ can be deduced and plotted in Fig.\ref{p1} (d). Each data point corresponds to a specific $\mathcal{L}$ value and is fitted by an exponential decaying scaling $J_{\max}(\mathcal{L}) = J_{c0} \exp({-\mathcal{L}/\xi_{\text{eff}}})$. And, we obtain
\begin{equation}\label{eq:Jc_scaling_fit}
J_{\max} \approx 0.955 \, \exp\bigl(-0.40\,\mathcal{L}\bigr).
\end{equation}
Therefore, the effective coherence length is roughly $\xi_{\text{eff}} \approx 2.5$ across the weak link.

\subsection{Dependence on the steepness $\sigma$}\label{subsec:steepness}

\begin{figure}[htbp]
\centering
\includegraphics[trim=0.cm 0.cm 0cm 0cm, clip=true, scale=0.45]{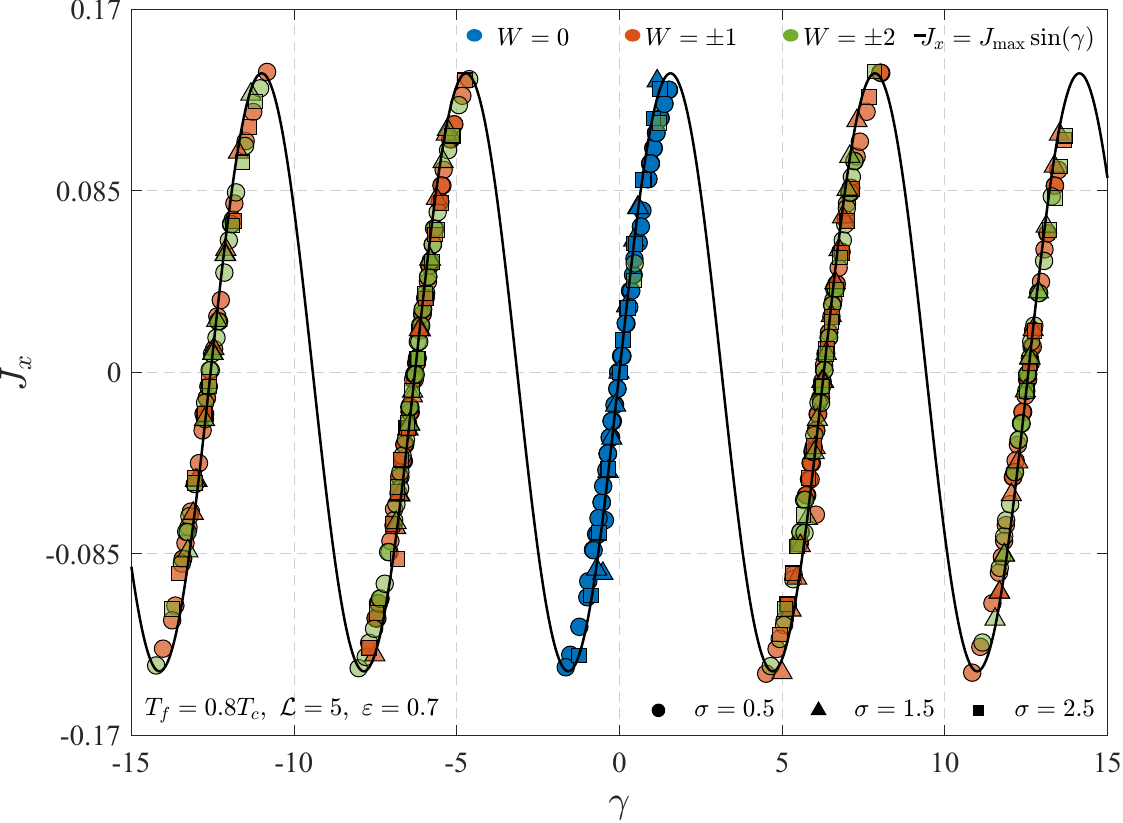}
\caption{Dependence of the current $J_x$ on the phase difference $\gamma$ across the weak link. The black solid curve represents the sinusoidal relation $J_x=J_{\text{max}}\sin (\gamma)$. Symbols indicate numerical results for different winding numbers $W$ (blue for $W=0$,  red for $W=\pm 1$, green for $W=\pm 2$) and varying steepnesses of the Josephson junction for $\sigma = \{$0.5(circles), 1.5 (triangles), 2.5 (squares)$\}$. Other parameters are fixed at $T_f=0.8T_c$, $\mathcal{L}=5$, $\epsilon=0.7$.  }
\label{p2}
\end{figure}

We now turn to the dependence of the current on the steepness $\sigma$, which controls the abruptness of the charge density drop at the boundaries of the weak link. 
Figure~\ref{p2} exhibits the current-phase relation for the fixed global parameters $T_f=0.8T_c$, $\mathcal{L}=5$, and $\epsilon=0.7$, by varying the interface steepness $\sigma$. In this figure, distinct winding number states are in different colors: $W=0$ (in blue), $|W|=1$ (in red), and $|W|=2$ (in green), representing post-quench topological configurations imprinted by the KZM. Junction steepness $\sigma$ is distinguished via geometric types: $\sigma=0.5$ (in dots), $\sigma=1.5$ (in triangles), and $\sigma=2.5$ (in squares). The black curve is the best fitted line of the datasets, which is consistent with the seminal relation \eqref{CPR}, 
\be J_x \approx 0.13\sin\gamma. \ee
From the Fig.\ref{p2} we see that all of the data collapse together onto the fitted curve, therefore, the steepness $\sigma$ has little impact on the current-phase relation. 

Besides, we also observe that for $W=0$ states, they mostly distribute near the vicinity of $\gamma=0$, while for $W\neq0$ they mostly distribute away from the vicinity of $\gamma=0$. In addition, the data will occupy in the increasing parts of the sinusoidal relation rather than locate in the decreasing parts of this relation. The reasons are similar presented in the preceding subsection.

\subsection{The effect of depth $\epsilon$}\label{subsec:depth}

 \begin{figure}[h]
\centering
\includegraphics[trim=0cm 0.cm 0cm 0cm, clip=true, scale=0.385]{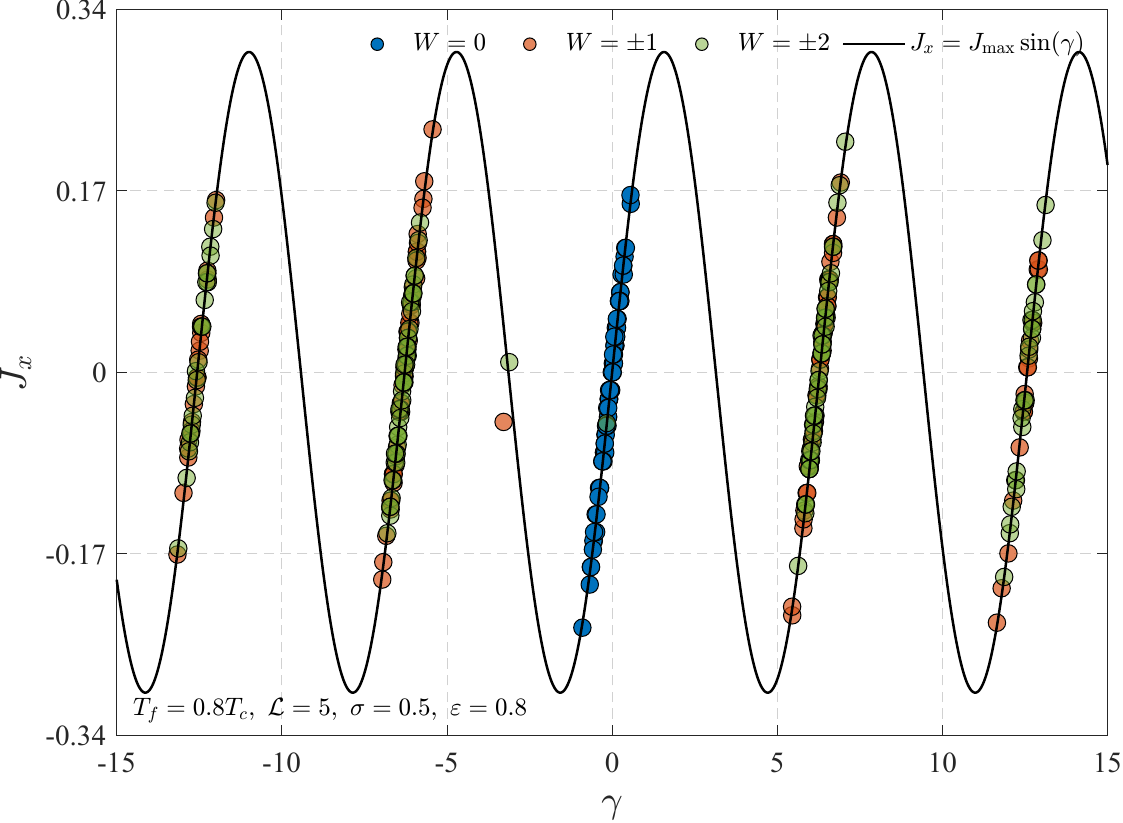}
\put(-215,150){(a)}~~~
\includegraphics[trim=0cm 0cm 0cm 0cm, clip=true, scale=0.385]{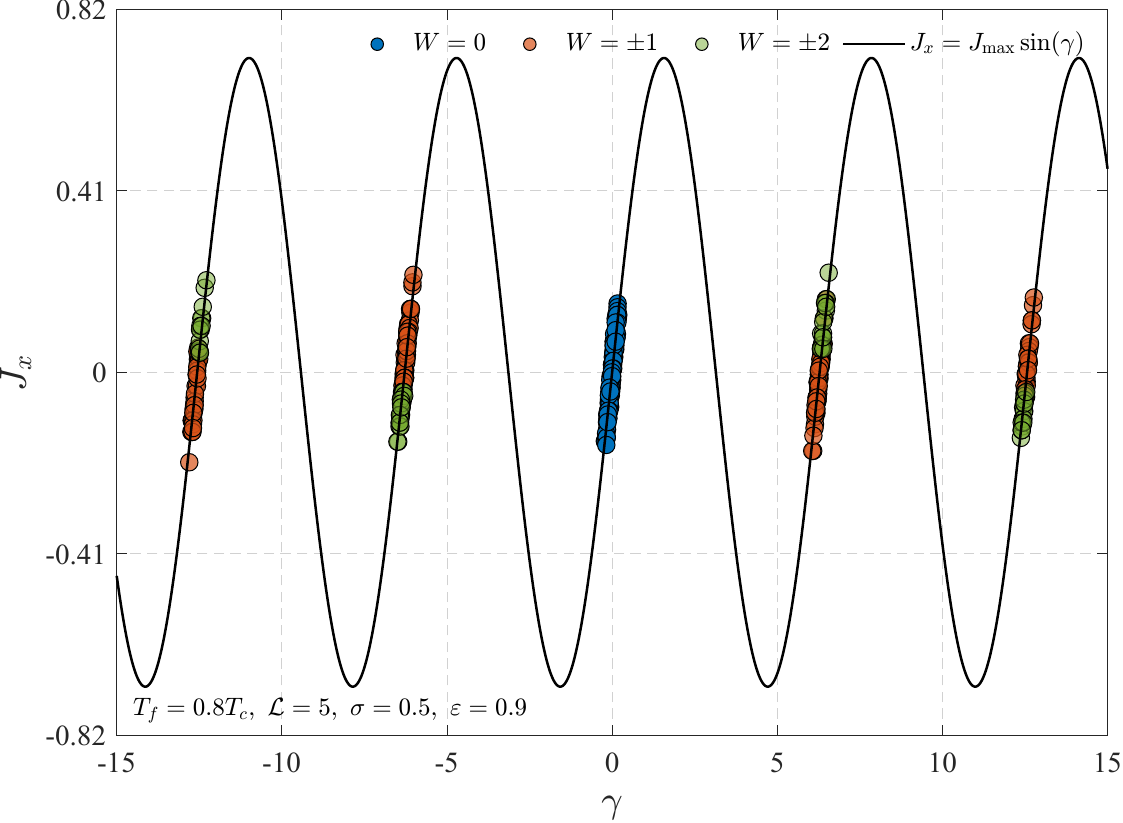}
\put(-215,150){(b)}\\
\includegraphics[trim=0.1cm 0.cm 0.cm -1cm, clip=true, scale=0.36]{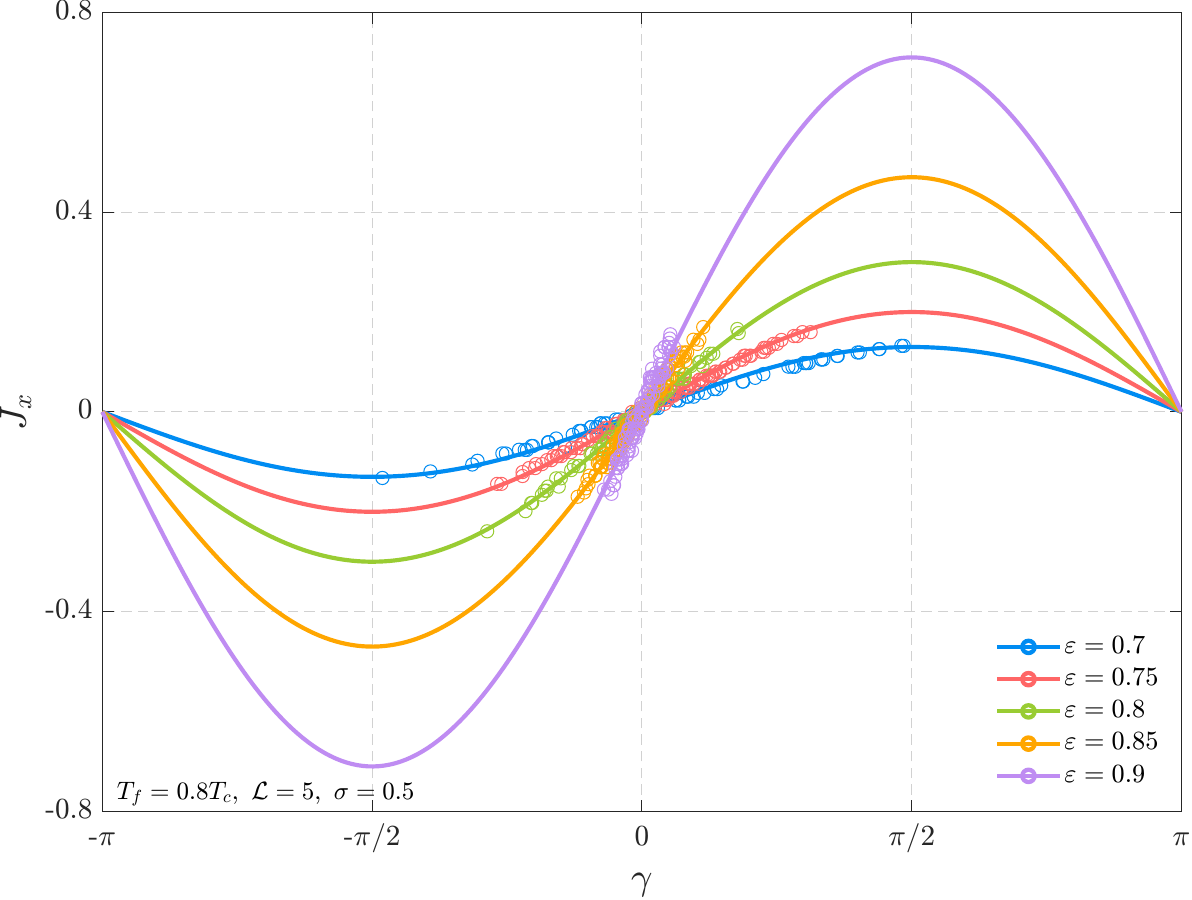}
\put(-215,150){(c)}~~~~
\includegraphics[trim=-0cm -0.5cm 0cm 0cm, clip=true, scale=0.38]{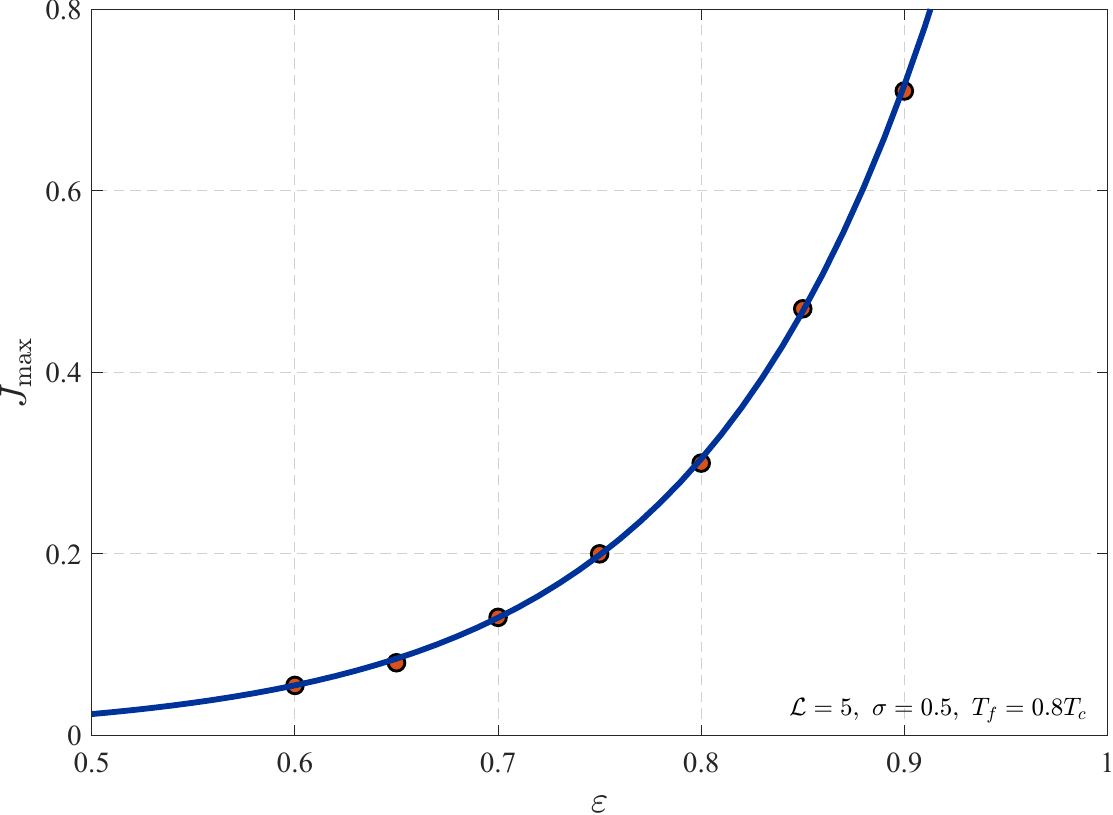}
\put(-215,150){(d)}
\put(-185,140){{\bf --} \footnotesize{$J_{\text{max}}\approx 3.263 \times 10^{-4} \, \exp\bigl(8.548\,\epsilon\bigr)$}}
\caption{Josephson current $J_x$ against the phase difference $\gamma$ with various junction depth $\epsilon$. (a, b) Current-phase relations for $\epsilon=0.8$ and $\epsilon=0.9$, respectively. Different winding numbers are denoted with colored circles while the solid curves are the best fit of the numerical data; (c) $J_x(\gamma)$ curves with increasing junction depth $\epsilon$, demonstrating the progressive modulation of the critical current. (d) Scaling of the critical supercurrent $J_{\text{max}}$ with $\epsilon$, revealing a monotonic enhancement of the Josephson critical current. Other parameters are fixed as: $T_f=0.8T_c$, $\mathcal{L}=5$, $\sigma=0.5$. }\label{p4}
\end{figure}

Now we are going to study the effect of the junction's depth $\epsilon$ on the current-phase relation.  
In Fig.~\ref{p4}(a) and \ref{p4}(b), we have fixed the parameters $T_f=0.8T_c$, $\mathcal{L}=5$, and $\sigma=0.5$ and varied the depth of the junction $\epsilon$. Both panels exhibit the supercurrent $J_x$ against the phase difference $\gamma$.
The numerical data correspond to different winding numbers $W$ with $W=0$ (in blue), $|W|=1$ (in red), $|W|=2$ (in green). The black solid curves are the best fit of those data and they satisfy the Josephson relation as $J_x = J_{\text{max}}\sin\gamma$.  
Panel (a) corresponds to a moderately deep potential ($\epsilon=0.8$), where the charge density suppression in the weak link reduces the local superfluid density, constraining the current-phase relation amplitude to $J_x \in [-0.34, 0.34]$; In contrast, panel (b) probes a shallower weak link ($\epsilon=0.9$), which elevates the superfluid stiffness in the constriction and raises the current-phase relation amplitude to $J_x \in [-0.82, 0.82]$. 
Therefore, we can conclude that the sinusoidal relation between the supercurrent and the phase difference persists even with different winding numbers. However, increasing the depth $\epsilon$ enhances the critical current without distorting the harmonic form of the Josephson relation, which is a stark contrast to the $\sigma$-independence discussed in the preceding subsection Sec.~\ref{subsec:steepness}. 

From Fig.\ref{p4}(a) we find similar distributions of the datasets as before. However, it is interesting to see a few exceptions: two points at around $\gamma\approx-\pi$ (in green and red) locate at the decreasing part of the sinusoidal line, which was never seen in previous subsections. These exceptions are expected from the stochastic features of the non-equilibrium dynamics of our model. However, as we explained before, from the point view of the free energy, these exceptional case are really rare.

Fig.\ref{p4} (c) examines the geometric scaling of the current-phase relation $J_x(\gamma)$ for the topologically trivial sector ($W=0$) as the depths of junction $\epsilon$ increases from 0.7 to 0.9. With the phase difference $\gamma$ scanned over $[-\pi, \pi]$, the panel (c) displays five curves from blue ($\epsilon=0.7$) to purple ($\epsilon=0.9$), which collectively illustrate the dependence of the Josephson current to $\epsilon$. In particular,  the extracted fitting parameters are summarized as follows:

\begin{subequations}
\begin{align}
J_x &\approx 0.13 \sin (\gamma), ~~~~~~\text{for}~~~ \epsilon = 0.7, \\
J_x &\approx 0.20 \sin (\gamma), ~~~~~~\text{for}~~~ \epsilon = 0.75, \\
J_x &\approx 0.30 \sin (\gamma), ~~~~~~\text{for}~~~ \epsilon = 0.8, \\
J_x &\approx 0.47 \sin (\gamma), ~~~~~~\text{for}~~~ \epsilon = 0.85,\\
J_x &\approx 0.71 \sin (\gamma), ~~~~~~\text{for}~~~ \epsilon = 0.9.
\end{align}\label{eq19}
\end{subequations}

From these Eqs.\eqref{eq19}, the persistence of the sinusoidal form confirms that $\epsilon$ modulates the amplitude of the Josephson current, leaving the harmonicity of the current-phase relation intact. Physically, the monotonic elevation of $J_{\text{max}}$ with $\epsilon$ reflects the progressive reinforcement of the order parameter amplitude within the weak-link.
The scaling of the critical current with junction depth is depicted in Fig.~\ref{p4}(d), and we find
\begin{equation}\label{eq:Jmax_epsilon_exp}
J_{\text{max}}(\epsilon) \approx 3.263 \times 10^{-4} \, \exp\bigl(8.548\,\epsilon\bigr).
\end{equation}
The extracted fit reveals a persistent exponential amplification of the current as $\epsilon$ increases, highlighting the highly nonlinear response of the holographic superfluid.

\subsection{The impact of final temperature $T_f$}\label{subsec:temperature}

\begin{figure}[htbp]
\centering
\includegraphics[trim=0cm 0cm 0cm 0cm, clip=true, scale=0.385]{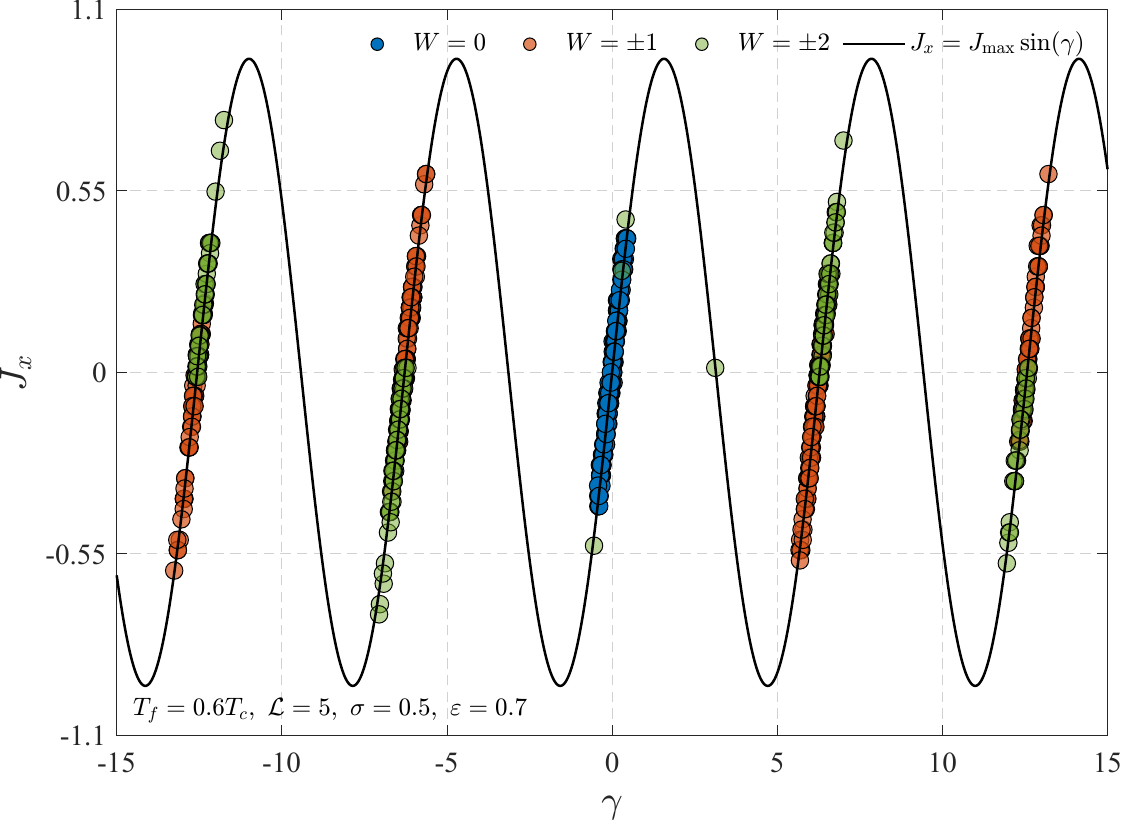}
\put(-213,150){(a)}~~~
\includegraphics[trim=0cm 0cm 0cm 0cm, clip=true, scale=0.385]{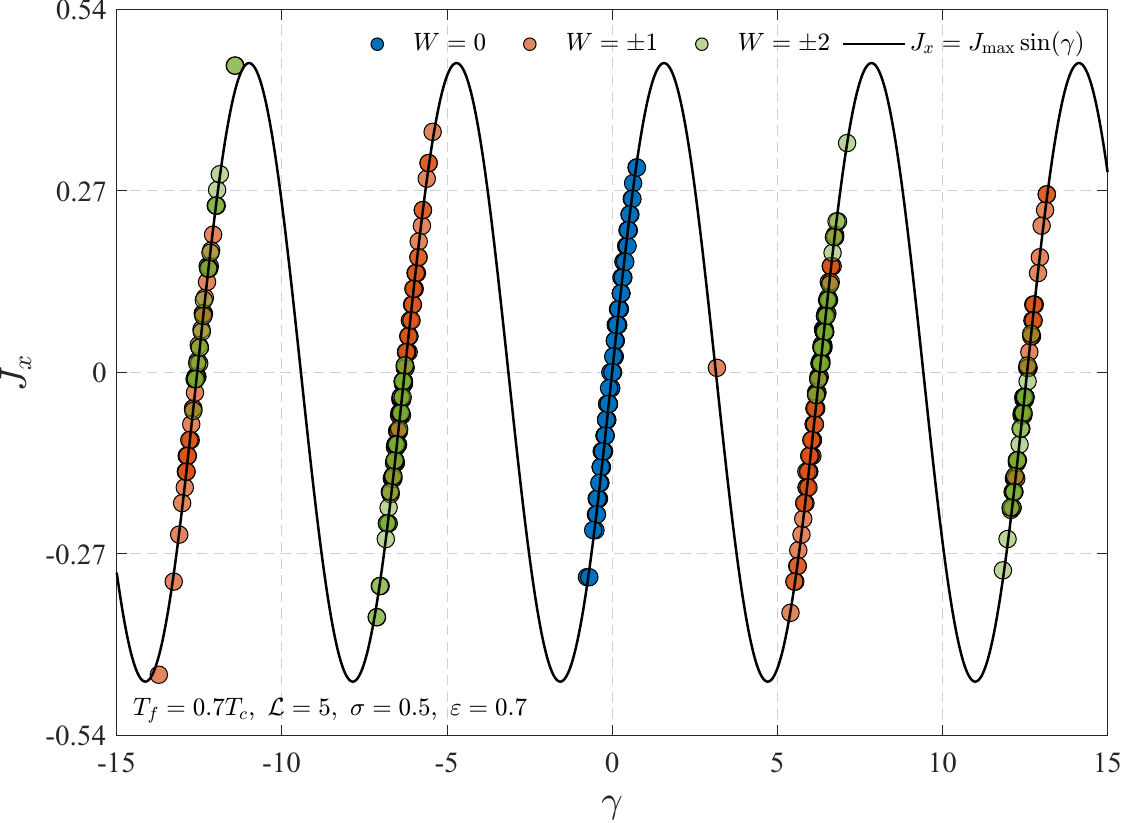}
\put(-213,150){(b)}\\
\includegraphics[trim=0cm 0cm 0cm -1cm, clip=true, scale=0.36]{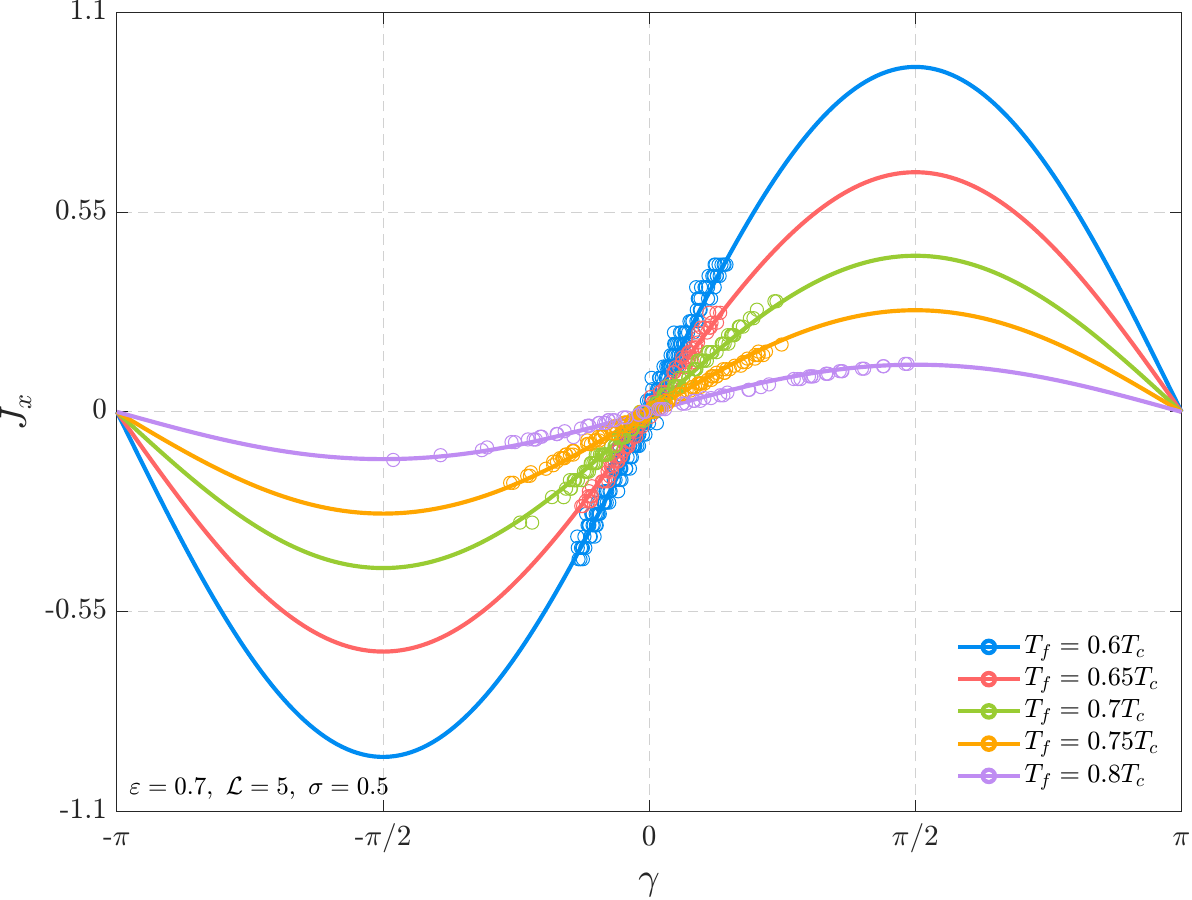}
\put(-215,150){(c)}~~~~
\includegraphics[trim=-0.5cm -0.1cm 0cm 0cm, clip=true, scale=0.38]{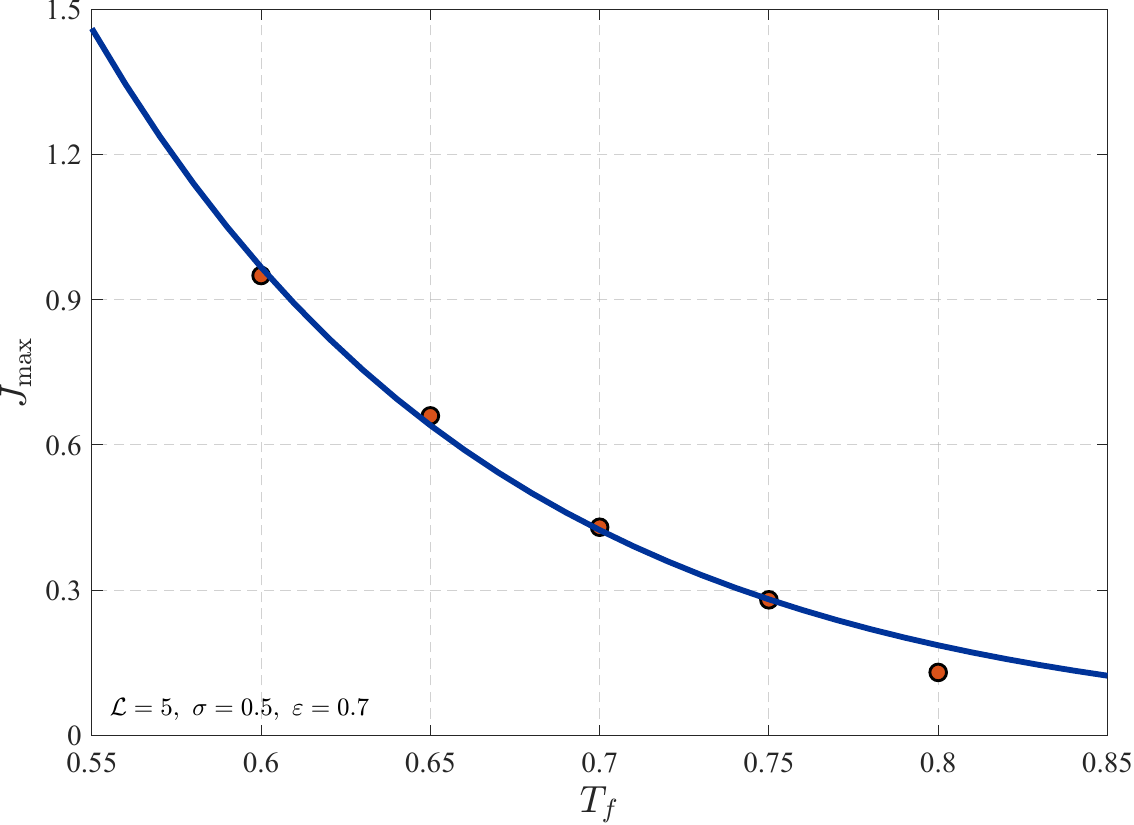}
\put(-215,150){(d)}
\put(-150,135){{\bf--} \footnotesize{$J_{\max}\approx 135.8 \, \exp{\bigl(-8.549\,T_f\bigr)}$}}
\caption{The current-phase relation $J_x(\gamma)$ is shown, with emphasis on how the final temperature $T_f$ reshapes this relation. Panels (a) and (b) display the current-phase relation at $T_f=0.6T_c$  and $T_f=0.7T_c$, respectively. The colored circles mark the quantized phase windings $W$, while solid curves are the sinusoidal fit of the data. In panel (c), successive $J_x(\gamma)$ curves are plotted with various $T_f$'s for $W=0$ case. Panel (d) plots the critical current $J_{\text{max}}$ against $T_f$, confirming an exponential decaying of the Josephson critical current. Other parameters are fixed as $\mathcal{L}=5$, $\sigma=0.5$ and $\epsilon=0.7$. }
\label{p5}
\end{figure}

Having established the geometric dependence of the critical current, we now turn to the thermodynamic variable governing the non-equilibrium phase transition: the final quench temperature $T_f$. In contrast to the structural parameters ($\mathcal{L}, \sigma, \epsilon$) of the junction, $T_f$ controls the thermal proximity to the critical point $T_c$ and regulates the magnitude of the local order parameter $\langle \mathcal{O} \rangle$ throughout the ring.

Figures~\ref{p5}(a) and \ref{p5}(b) show the influence of the final quench temperature $T_f$ on the current-phase relations, by fixing other parameters as $\mathcal{L}=5$, $\sigma=0.5$ and $\epsilon=0.7$. In both panels the colored points correspond to different winding numbers $W$, such that $W=0$ (in blue), $|W|=1$ (in red) and $|W|=2$ (in green). The black solid line is the best fit of the numerical data, having the universal sinusoidal relation $J_x = J_{\max}\sin\gamma$. From these two panels we can see that most of the data points are distributed similarly as those in previous subsections. However, there are still some exceptions as we have discussed in Fig.\ref{p4}(a). In Fig. \ref{p5}(a) and (b), there are points at around $\gamma=\pi$ (in green and red) which locate at the decreasing parts of the sinusoidal relations. The reasons are similarly due to the stochastic properties of the non-equilibrium dynamics of our model.

Fig.~\ref{p5}(c) exhibits the current-phase relation in the topologically trivial sector ($W=0$) with five discrete temperatures $T_f/T_c \in \{0.6, 0.65, 0.7, 0.75, 0.8\}$. The sinusoidal forms confirm that $T_f$ modulates  the amplitude of the Josephson current without perturbing the harmonicity of the phase relation, which are fitted as follows,
\begin{subequations}
\begin{align}
J_x &\approx 0.95 \sin (\gamma), ~~~~~~\text{for}~~~ T_f = 0.6T_c, \\
J_x &\approx 0.66 \sin (\gamma), ~~~~~~\text{for}~~~ T_f = 0.65T_c, \\
J_x &\approx 0.43 \sin (\gamma), ~~~~~~\text{for}~~~ T_f = 0.7T_c, \\
J_x &\approx 0.28 \sin (\gamma), ~~~~~~\text{for}~~~ T_f = 0.75T_c,\\
J_x &\approx 0.13 \sin (\gamma), ~~~~~~\text{for}~~~ T_f = 0.8T_c.
\end{align}
\end{subequations}
It is seen that the current-phase relation maintains a sinusoidal profile, however, its magnitude $J_{\text{max}}$ exhibits a monotonic suppression by increasing the final temperature $T_f$, reflecting the thermal degradation of the condensate.
The scaling of the critical current $J_{\text{max}}$ with temperature is plotted in Fig.~\ref{p5}(d). They satisfy a pronounced exponential decaying as,
\begin{equation}\label{eq:Jmax_Tf_exp}
J_{\max}(T_f) \approx 135.8 \, \exp\bigl(-8.549\,T_f\bigr).
\end{equation}
This formula is consistent with previous holographic studies in \cite{Horowitz:2011dz,Wang:2011ri,Wang:2012yj}.

\section{Conclusions and discussions}\label{four}
In this work, we have performed a comprehensive investigation on the non-equilibrium dynamics and the current-phase relations of a holographic Josephson junction embedded within a superfluid ring. By leveraging the KZM to generate topological phase windings and utilizing a spatially modulated charge density $\rho(x)$, we successfully obtained the seminal sinusoidal relation of the current-phase relation of a Josephson junction.

After the relaxation of non-equilibrium dynamics, the current-phase relations will exhibit periodic profiles due to the compact geometry of the ring. For the $W=0$ case, they mostly distribute near $\gamma=0$ phase difference. However, for the $|W|=\pm1$ and $|W|=\pm2$ cases, they will mostly scatter away from $\gamma=0$ phase differences. Besides, the sinusoidal current-phase relation will only exhibit one-half side, i.e., they will occupy the increasing part of the sinusoidal relation, rather than the decreasing part of this relation. We attribute this phenomenon to the requirement of the lower free energy of the system. However, there is still very rare chance for them to distribute in the decreasing part of the sinusoidal relation, which is due to the stochastic properties of the non-equilibrium dynamics as the initial conditions of the system.

Furthermore, our analysis of the current-phase relation revealed a rich phenomenon governed by four distinct control parameters: First, we investigated the relationship between the junction width $\mathcal{L}$ and the current-phase relation, which is characterized by an exponential decay of the critical current $J_{\text max}$; Second, by tuning the steepness of the junction $\sigma$, we found that the steepness of the junction had little impact on the current-phase relation;  Third, the depth $\epsilon$ of the junction had an exponential growing relation to the critical current; Finally, we elucidated the thermodynamic dependence of the Josephson current-phase relation. While the scaling $J_x \propto \sin\gamma$ persists across all probed temperatures, the critical current amplitude undergoes exponential suppression with the increasing $T_f$. 
The consistency of the sinusoidal current-phase relation across all parameters emphasizes the robustness of the holographic superconducting state, while the distinct responses of the critical current $J_{\text{max}}$ to the parameters $\mathcal{L}$, $\epsilon$, and $T_f$ provide a tunable framework for engineering Josephson devices. Notably, the exponential sensitivity of $J_{\text{max}}$ to $\epsilon$ and its thermal attenuation offer experimentally testable predictions for condensed matter realizations of holographic models. 

It is worth noting that our dynamical method is more realistic and universal than previous static methods. The novelty and importance lie in: First, the non-equilibrium dynamics in our study provides a test bed to mimic the real experiments in preparing the superconducting states, since in real experiments people also need to quench the system into a superconducting state rather than get a superconducting material in the very beginning;  Second, the dynamical case in our study offers a truly visible phase difference across the junction by comparing to the static case. This is because in the previous static studies, people used the gauge-invariant quantities by gauging away the phase of the order parameter \cite{Horowitz:2011dz,Wang:2011rva,Wang:2011ri}. Thus, there is no any phase of the order parameter in the equations of motion. However, in our dynamical case we used the complex scalar field and the phase difference is vividly shown in Fig.\ref{p0}. One can read off the phase difference directly across the junction. This is a big advantage of the dynamical study compared to the static case; Third, our study reveals a profound truth that the stochastic properties of the non-equilibrium dynamics combining with the relaxation to the final equilibrium state can perfectly simulate the previous studies with only static quantities.  Our method is universal and can be readily extended to other models, such as in studying the holographic Little-Parks effects in \cite{Li:2021mtd}.

\section*{Acknowledgements}

This work was partially supported by the National Natural Science Foundation of China (Grants No.12075143, No.12175008, No.12305067 and No.12675057), Natural Science Foundation of Shanxi Province, China (Grant No.2025030221211241) and Shanxi Provincial Youth Scientific Research Project (Grant No. 202303021222209 ).

\end{document}